\documentclass[12pt]{article}

\begin{document}

\begin{center}
{\Large\bf On source coupling and the teleparallel equivalent to GR}
\end{center}

\begin{center}
Lau Loi So\footnote{e-mail address: s0242010@cc.ncu.edu.tw} and
James
M. Nester\footnote{e-mail address: nester@phy.ncu.edu.tw}\ \\
$^{1}$Department of Physics, National Central University,
Chung-Li 32054, Taiwan.\\
$^{2}$Department of Physics and Institute of Astronomy, National
Central University, Chung-Li 32054, Taiwan.
\end{center}



\begin{abstract}
Alternatives to the usual general relativity (GR) Riemannian
framework include Riemann-Cartan and teleparallel geometry.  The
``teleparallel equivalent of GR" (TEGR, aka GR${}_{||}$) has certain
virtues, however there have been allegations of serious source
coupling limitations. Now it is quite straightforward to show that
the coupled dynamical field equations of Einstein's GR with any
source can be accurately represented in terms of any other
connection, in particular teleparallel geometry. Using an argument
similar to one used long ago to show the ``effective equivalence"
between GR and the Einstein-Cartan theory, we construct the
teleparallel action which is equivalent to a given Riemanian one;
thereby finding the ``effectively equivalent" coupling principle for
all sources, including spinors. No auxiliary field is required.  Can
one decide which is the real ``physical" geometry?  Invoking the
minimal coupling principle may give a unique answer.
\end{abstract}

\section{Background}
The world has geometry, but which kind?\cite{at}  In general,
geometry includes both a metric and a connection. They determine
three tensors: the non-metricity $Q$, torsion $T$ and curvature
$R$. Here we consider only the vanishing $Q$ cases: Riemann-Cartan
($R\ne0\ne T$), Riemannian ($R\ne0=T$) and  teleparallel ($R=0\ne
T$).

Aldrovandi and Pereira\cite{AP} have emphasized that spaces do not
have curvature and torsion, rather it is connections that have
curvature and torsion.  Since any two connections differ by a
tensor, it is straightforward to change from one connection to any
other. In this way we can take the field equations of, for
example, the Poincar{\'e} Gauge theory and represent them in terms
of the Riemannian connection. The price is that one needs to add
to the Riemannian geometry an extra tensor field (e.g., the
contortion, or torsion). And, moreover, the extra field couples in
a characteristic ``non-minimal" fashion. As long as one allows an
extra tensor field with its associated special type of non-minimal
coupling, one can transcribe any physical equations from
Riemann-Cartan to Riemannian to teleparallel geometry and vice
versa.
 For a complete ``equivalence'' we also want the conserved
quantities, like energy-momentum and angular momentum, hence we
should also find the equivalent action. The effectively equivalent
action also generally needs extra fields and non-minimal coupling.

There is a teleparallel equivalent to GR (TEGR, aka
GR${}_{||}$).\cite{Mol,Mal,MGH,AGP} The simplest description is in
terms of orthonormal-teleparallel (OT) frames. It is easily
constructed: given the GR metric, simply choose any orthonormal
frame field and declare it to be parallel (i.e., take the
connection coefficients to vanish in this frame).

It has been alleged that GR${}_{||}$ has serious source coupling
limitations, e.g. it has been said that only scalar matter fields
or gauge fields are allowed as sources, whereas matter carrying
spin cannot be consistently coupled;\cite{Gron} some argue that
torsion does not couple to electromagnetism,\cite{em} others have
a different opinion.\cite{tem}

For the special case of GR and GR${}_{||}$ we can use an argument
similar to that used long ago to establish the effective
equivalence of GR and the Einstein-Cartan (EC) theory\cite{eq}. As
in that case, we find that the GR${}_{||}$ effective equivalent
field equations and action also do not require an extra field.
Nevertheless we get a complete equivalence for all sources,
including, in particular, spinors.

\section{Riemannian equations re-expressed in teleparallel form}
The Riemannian GR equations for any source follow from an action
of the form:
\begin{equation}
L=ee_{a}{}^{\mu}e_{b}{}^{\nu}R^{ab}{}_{\mu\nu}(\Gamma)
  +T^{a}{}_{\mu\nu}\tau_{a}{}^{\mu\nu}+L(\phi,e,\partial\phi+\Gamma\phi)
\end{equation}
by variation with respect to the frame $e^a{}_\mu$, the connection
$\Gamma$, the source field $\phi$ and the multiplier $\tau$ (which
enforces the vanishing of torsion for $\Gamma$).

These standard GR dynamical equations can be represented in terms
of any other connection, e.g. Riemann-Cartan with torsion, simply
by a transformation involving an extra field:
$\Gamma^{a}{}_{b\mu}={\bar\Gamma^{a}{}_{b\mu}}+K^{a}{} _{b\mu}$,
here $K^{a}{}_{b\mu}$ is the contorsion tensor (linearly related
to the torsion). Minimally coupled Riemannian equations become
non-minimally coupled to the contorsion.  In particular we can
also force the new connection ${\bar\Gamma}$ to be teleparallel,
so that ${\bar R}{}^{ab}{}_{\mu\nu}=0$.

Within each of the dynamic equations we make the above
substitution. In this way, all of these Riemannian equations of
motion can be transcribed into teleparallel form. But the form of
the coupling has become non-minimal. What about the conserved
quantities?  Can we find an action that gives the new equations?

\section{Equivalent Teleparallel Action}
For complete equivalence we need an equivalent teleparallel
action.  It can be obtained by the same substitution
$\Gamma^{a}{}_{b\mu}={\bar\Gamma}{}^{a}{}_{b\mu}+K^{a}{}_{b\mu}$.
Expanding the Hilbert scalar curvature Lagrangian of GR gives
$\bar R + \bar D K +K^2$.  We can enforce vanishing teleparallel
curvature with a Lagrange multiplier field.  After removing a
total divergence the effectively equivalent action is of the form
\begin{equation}
L_{||}=-\frac{1}{4}{\bar T}^{\alpha\mu\nu}{\bar T}_{\alpha\mu\nu}
-\frac{1}{2}{\bar T}^{\alpha\beta\nu}{\bar T}_{\alpha\beta\nu}
+{\bar T}^{\alpha}{}_{\alpha\mu} {\bar T}_{\beta}{}^{\beta\mu} +
{\bar R}^{ab}{}_{\mu\nu}\lambda_{ab}{}^{\mu\nu}
+L(\phi,e,\bar\nabla\phi,K),
\end{equation}
to be varied with respect to the source $\phi$, the frame
$e^a{}_\mu$, the teleparallel connection $\bar\Gamma$, and the
multiplier $\lambda$. In short:
\begin{eqnarray}
\begin{array}{ccc}
L_{GR}& \stackrel{\delta}{\longrightarrow}&\mbox{Riemannian form of equations}\\
\downarrow & & \downarrow\\
L_{||}&\stackrel{\delta}{\longrightarrow}&\mbox{teleparallel form of equations}\\
\end{array} \nonumber
\end{eqnarray}
where $\downarrow$ is $\Gamma\to\bar{\Gamma}+K$. As in the EC case
studied earlier, (unlike the corresponding situation for the
Poincar{\'e} Gauge theory or Metric-Affine gravity) we do not need
any extra field. We obtain a complete equivalence for all sources
--- including spinors.

\section{Conclusion}
There is a one-to-one correspondence between the two
representations.  No physical experiment can distinguish between
them; no observation can decide whether spacetime has curvature
without torsion or torsion with vanishing curvature. GR and
GR${}_{||}$ are completely equivalent for all sources.  TEGR
truely is a good name.

Can we determine the true ``physical" geometry?  Is it Riemannian
or teleparallel?  This may be possible with the aid of the minimal
coupling principle.  Sources which are minimally coupled with
respect to the Riemannian geometry will not generally be minimally
coupled with respect to the teleparallel geometry, and vice versa.
We could hope that observations would yield results which are
consistent with one of these geometries and the associated minimal
coupling.

\section*{Acknowledgement} This work was supported by the National
Science Council of the ROC under grant no NSC 92-2112-M-008-050.

\end{document}